%% file: NISN_junction.tex
\documentclass[aps,pra,superscriptaddress,twocolumn,10pt]{revtex4-1}

\pdfoutput=1

\usepackage{amsmath}
\usepackage{latexsym}
\usepackage{graphicx}
\usepackage{amsfonts}
\usepackage{braket}
\usepackage{xcolor}
\definecolor{darkblue}{rgb}{0,0,0.5} 
\usepackage[colorlinks=true,
        urlcolor=darkblue,
        anchorcolor=darkblue,
        linkcolor=darkblue,
        citecolor=darkblue,
        pdfauthor={Johan B. C. Engelen},
        pdfkeywords={SVG; LaTeX; Inkscape},
        pdftitle={How to include an SVG image in LaTeX},
        pdfsubject={Describes how to include an SVG image easily in LaTeX using Inkscape}] {hyperref}
\usepackage{natbib}
\usepackage{amssymb}
\usepackage{nicefrac}
\usepackage{fancyhdr}
\usepackage[utf8]{inputenc}
\usepackage{dsfont}
\usepackage{colortbl}
\usepackage{transparent}
\usepackage{url}
\usepackage[protrusion=true,expansion=true]{microtype}
\usepackage{subfigure}
\usepackage{psfrag}
\usepackage{tikz}
\usepackage{paralist}
\usepackage{ulem}
\usepackage[section]{placeins}
\usepackage{mathtools}
\usepackage{balance}
\usepackage{lastpage}
\mathtoolsset{showonlyrefs}
\usepackage{fourier}
\usepackage{placeins}
\usepackage{calc}
\usepackage{pgf}

\usepackage{verbatim}
\makeatletter
\g@addto@macro\@verbatim\small
\makeatother 

\newcommand{\executeiffilenewer}[3]{%
 \ifnum\pdfstrcmp{\pdffilemoddate{#1}}%
 {\pdffilemoddate{#2}}>0%
 {\immediate\write18{#3}}\fi%
}

\newcommand{\spd}[3]{\ifthenelse{\equal{#2}{#3}}{\frac{\partial^2 #1}{\partial {#2}^2}}{\frac{\partial^2 #1}{\partial #2 \partial #3}}}
\newcommand{\cc}[1]{\ensuremath{#1^*}}
\newcommand{\tr}[1]{\ensuremath{\text{Tr}\left\{ #1 \right\}}}
\newcommand{\abs}[1]{\ensuremath{\left| #1 \right|}}
\newcommand{\comm}[2]{\ensuremath{\left[ #1, #2 \right]}}

\newcommand{\Tr}[1]{\ensuremath{\text{Tr}\left\{ #1 \right\}}}
\newcommand{\normalcurrent}{normal current}
\newcommand{\fig}{Fig.}
\newcommand{\Fig}{Fig.}
\newcommand{\eq}{Eq.}

\newcommand{\D}{\ensuremath{\Delta_{\text{BCS}}}}
\newcommand{\n}{\ensuremath{n_{\text{qp}}}}
\newcommand{\e}{\ensuremath{\epsilon}}
\newcommand{\SWOT}{setup 1}
\newcommand{\SWT}{setup 2}
\newcommand{\de}{\ensuremath{\Delta}}
\newcommand{\sect}{Sec.}

\renewcommand{\d}[1]{\ensuremath{\mathrm{d}#1}}
\renewcommand{\cos}[1]{\ensuremath{\text{cos}\left( #1 \right)}}
\renewcommand{\sin}[1]{\ensuremath{\text{sin}\left( #1 \right)}}

\renewcommand{\Re}[1]{\ensuremath{\text{Re}\left\{ #1 \right\}}}
\renewcommand{\Im}[1]{\ensuremath{\text{Im}\left\{ #1 \right\}}}
\renewcommand{\ln}[1]{\ensuremath{\text{ln} \left( #1 \right)}}

\renewcommand{\cosh}[1]{\ensuremath{\text{cosh}\left( #1 \right)}}

\renewcommand{\exp}[1]{\ensuremath{{\rm e}^{#1}}}

\usetikzlibrary{arrows,shapes,decorations.pathmorphing}

\begin{abstract}
The performance of many superconducting devices is degraded in presence of non-equilibrium quasiparticles in the superconducting part. One promising approach towards their evacuation is the use of normal-metal quasiparticle traps, where normal metal is brought into good metallic contact with the superconductor. A voltage biased normal-metal--insulator--superconductor junction equipped with such a trap is used to investigate on the trapping performance and the part played by the superconducting proximity effect therein. This involves an appropriate one-dimensional model of the junction and the numerical solution of Usadel equations describing the non-equilibrium state of the superconductor. The functionality of the trap is determined by the density of states (DOS) at the tunnel barrier. Herein, the proximity effect leads to two antagonistic characteristics affecting the trapping performance: the beneficial reduction of the DOS at an energy $|E| = \D$ versus the contraction of the spectral energy gap causing quasiparticle poisoning. For both effects the trap position is decisive, which needs to be taken into account for optimizing the trapping performance. In addition, the conversion between dissipative normal and supercurrent inside the superconducting part with its impact on the quasiparticle density is studied.
\end{abstract}

\makeindex

\begin{document}

\title{Role of the proximity effect for normal-metal quasiparticle traps}

\author{R. P. Schmit}
\affiliation{Theoretical Physics, Saarland University, Campus, 66123 Saarbr{\"u}cken, Germany}
\author{F. K. Wilhelm}
\affiliation{Theoretical Physics, Saarland University, Campus, 66123
Saarbr{\"u}cken, Germany}

\maketitle

\section{Introduction}
\label{sec:intro}
Mesoscopic superconductors are easily driven out of equilibrium, often leading to the generation of quasiparticles (QPs). Furthermore, there is convincing experimental evidence for the existence of a residual QP population even at low temperatures~\cite{shaw2008kinetics, riste2013millisecond, stern2014flux, de2011number, paik2011observation}, exceeding the expected equilibrium density. The non-equilibrium QPs have a detrimental impact on most superconducting devices, e.g. causing decoherence in superconducting qubit systems~\cite{paik2011observation, catelani2011relaxation, catelani2012decoherence, catelani2014parity, lutchyn2005quasiparticle, lutchyn2006kinetics, leppakangas2012fragility, martinis2009energy}, lowering the efficiency of micro-refrigerators~\cite{pekola2000trapping, giazotto2006opportunities, rajauria2007electron, muhonen2012micrometre}, or preventing the experimental detection of the 2e periodic Coulomb staircase in single Cooper pair transitors~\cite{joyez1994observation, lehnert2003measurement, aumentado2004nonequilibrium, gunnarsson2004tunability}.
 
Sufficient cooling down to temperatures far below the critical temperature might help for some technical applications since thermal QPs are (almost) absent due to the spectral energy gap in the excitation spectrum. One important process during QP relaxation is their electron-phonon mediated recombination~\cite{kaplan1976quasiparticle} to form Cooper pairs, along with the emission of phonons with energy $\hbar \omega \gtrsim 2 \Delta$. The related time scale is controlled not only by the phononic DOS at $\hbar \omega \gtrsim  2 \Delta$, but also by the phonon's pair breaking potential to excite new QPs, effectively increasing their lifetime~\cite{levine1968recombination, rothwarf1967measurement, kaplan1976quasiparticle, patel2017phonon, catelani2010effect, owen1972superconducting}. Thus, reaching complete thermalization might take too long to be practical for most quantum computing applications based on superconducting elements. Furthermore, the generation of non-equilibrium QPs is intrinsic to qubit control techniques using single flux quantum pulse sequences~\cite{mcdermott2014accurate, liebermann2016optimal, leonard2019digital}, while different strategies to minimize QP generation and poisoning are available~\cite{leonard2019digital}.

\begin{figure} 
 \centering
  \def\svgwidth{1.1\columnwidth}
 \executeiffilenewer{setup3.svg}{setup3.pdf}%
 {inkscape -z -D --file=setup3.svg %
 --export-pdf=setup3.pdf --export-latex}%
 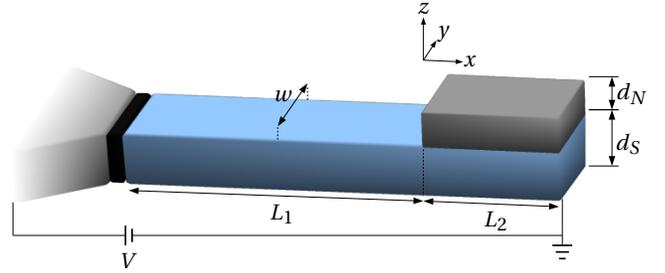%

  \caption{(Not to scale) Schematic of a QP injector with attached normal-metal trap. A mesoscopic superconducting wire (blue part) is connected to a big normal metal reservoir (partially shown by left gray part) via a thin tunnel barrier (black part). The reservoir is kept at temperature $T=0$ K and potential $eV$ measured from the superconductor's one. At its end the superconductor is covered in good metallic contact with another normal metal with length $L_2$, serving as QP trap, and is electrically grounded. In what follows, this setup is referred to as \SWT. The trapping performance is investigated by comparing the non-equilibrium steady states of this setup and another QP injector with the same dimensions and parameters but without attached trap. This reference setup is called \SWOT~in the following. The origin $x=0$ is located at the left edge of the QP trap. The dimensions $w$ and $d_i$ in the transverse directions $y$ and $z$ are assumed to be much smaller than the superconducting coherence length $\xi_0$, which allows for a one-dimensional model of the setup.} 
  \label{fig:setup} 
\end{figure}

Evacuating and trapping QPs in less active regions of the device seems to provide a practicable way to improve the device performance. Most of the current trapping techniques share a common principle: Spatial variations in the superconducting order parameter deform the energy landscape the QPs reside in, thereby introducing accumulation regions for the QPs where they are trapped in after relaxing. This can be achieved by engineering gap inhomogeneities~\cite{aumentado2004nonequilibrium, friedrich1997experimental, ferguson2008quantitative} directly affecting the order parameter, or by exploiting the superconducting proximity effect of a normal metal on a superconductor, which occurs in normal-metal vortex penetration due to external magnetic fields~\cite{ullom1998magnetic, peltonen2011magnetic, nsanzineza2014trapping, wang2014measurement, taupin2016tunable} or when purposely bringing both metals in good metallic contact~\cite{goldie1990quasiparticle, pekola2000trapping, ullom2000measurements, rajauria2012efficiency, knowles2012probing, nguyen2013trapping, saira2012vanishing, hosseinkhani2017optimal} \footnote{Exploiting the mutual influence of two superconductors with different bulk energy gaps has a similar effect~\cite{saira2012vanishing}.}. 

We focus on the QP trapping performance of the latter technique, a normal metal in good metallic contact with a superconductor. As is shown in \fig~\ref{fig:setup}, a superconducting island $S$ of length $L_S=L_1+L_2$ is connected via a tunnel junction ($I$) to a normal metal ($N$) reservoir hold at temperature $T = 0$ and potential $eV$ measured from the superconductor's one. QPs are injected into, diffuse through and exit the superconductor via the electrical grounding at the superconductor's end. A thin normal metal partially covering the superconducting island in a distance of $L_1$ to the injector serves as QP trap. 

We study the role of the proximity effect on the non-equilibrium steady-state of the superconducting island, especially in regard to the density of QPs. In order to determine the efficiency of the trap and its influence on the QP distribution $\n(x)$ inside the $S$-island, $\n(x)$ has to be compared with that of an NIS-junction with the same geometry but without a covering metal. In the following, we will refer to this reference setup as \SWOT, whereas the setup with normal-metal trap is referred to as \SWT.

Usually, the dynamics and the steady-state of the QP distribution are studied by using a phenomenological diffusion equation for the QP density, taking into account their interaction with phonons and loss mechanisms due to QP recombination and trapping~\cite{rothwarf1967measurement, chang1977kinetic, rajauria2009quasiparticle, rajauria2012efficiency, hosseinkhani2018proximity, wang2014measurement, patel2017phonon, lenander2011measurement, riwar2016normal}. Here, we follow a different approach and make use of the Usadel formalism~\cite{usadel1970generalized, belzig1999quasiclassical, volkov1998methods}, which is a convenient tool to study dirty mesoscopic proximity systems in and out of equilibrium and was applied in various fields~\cite{hosseinkhani2018proximity, gueron1996superconducting, anthore2003density, volkov1998proximity, seviour1998conductance, charlat1996reentrance, charlat1996resistive, belzig1996local, zaikin1996coherent, golubov1997coherent, virtanen2010theory, cuevas2006proximity, virtanen2007thermoelectric, kauppila2013nonequilibrium, voutilainen2005nonequilibrium}. In particular, it was used for a detailed theory of non-equilibrium phenomena in a superconductor in contact with a normal-metal trap, given in~\cite{voutilainen2005nonequilibrium}. This formalism gives access to spectral quantities such as the QP DOS, and also to non-equilibrium quantities such as the QP population and current densities.

The paper is structured as follows: In \sect~\ref{sec:Framework} we give the appropriate Usadel equations and matching conditions necessary for an inhomogeneous overlap geometry. The numerical results are presented in \sect~\ref{sec:NumRes}. We conclude with \sect~\ref{sec:conclusion} giving final remarks.

\section{Theoretical Framework}
\label{sec:Framework}

For mesoscopic systems in and out of equilibrium, the physical information can be encoded in the Keldysh Green's functions $\check{G}=\begin{pmatrix} \hat{R} & \hat{K} \\ 0 & \hat{A} \end{pmatrix}$, where $\hat{R}, \hat{A}$ and $\hat{K}$ are referred to as retarded, advanced and Keldysh component, respectively. In the dirty limit, i.e. when the QPs undergo frequent elastic scattering with a scattering rate $\tau^{-1}$ and their motion is diffusive due to a high impurity concentration, the Keldysh Green's functions obey the Usadel equations~\cite{usadel1970generalized}. In the case of superconductivity, the form of the underlying equations is the same as for normal metals when passing to the Nambu (or particle-hole) space~\cite{nambu1960quasi}. In this case, the components $\hat{R}, \hat{A}, \hat{K}$ of the Keldysh Green's functions become $2\times 2$-matrices, and the Usadel equations read 
\begin{align}
 \hbar D \nabla \left( \check{G} \nabla \check{G} \right) = \comm{-i E \check{\tau}_3 + \check{\Delta}}{\check{G}}.
 \label{eq:Usadel_Keld}
\end{align}
Here, $D = v_F^2\tau / 3$ with the Fermi velocity $v_F$ denotes the diffusion constant and 
\begin{align*}
 \check{\tau}_3=\begin{pmatrix} \hat{\tau}_3 & 0 \\ 0 & \hat{\tau}_3 \end{pmatrix}, ~~ \check{\Delta}=\begin{pmatrix} \hat{\Delta} & 0 \\ 0 & \hat{\Delta} \end{pmatrix}, ~~ \hat{\Delta} = \begin{pmatrix} 0 & \Delta \\ \cc{\Delta} & 0 \end{pmatrix}
\end{align*} 
with the Pauli matrices $\hat{\tau}_i$.

When the (inverse) proximity effect cannot be neglected, as it is the case for \SWT, these equations must be supplemented by the self-consistency equation for the order parameter,
\begin{align}
 \Delta=\frac{N_0 \lambda}{8 i} \int \limits_{-\hbar \omega_D}^{\hbar \omega_D} \Tr{\left( \hat{\tau}_1 - i \hat{\tau}_2 \right) \hat{K}}\d E. \label{eq:Delta_Keld}
\end{align}
Here, $\lambda$, $N_0$ and $\omega_D$ denote the strength of the attractive pair interaction, the normal-state DOS at the Fermi energy and the Debye frequency, respectively. 

The $N$-metal reservoir at $x=-L_1$ is assumed to be unaffected by the proximity effect. 
\subsection*{One-dimensional model}
Mesoscopic wires with transversal dimensions much smaller than the superconducting coherence length can be sufficiently treated as one-dimensional. In this section we show how the inhomogeneous overlap geometry shown in \fig~\ref{fig:setup} due to the only partial covering by the normal-metal trap can be approximated by a one-dimensional setup. This is done in two steps: First, we consider the homogeneous overlap part extending in the interval $x \in [0, L_2]$ ignoring the rest and derive one-dimensional Usadel equations for the Keldysh Green's function $\check{G}$. Essentially, they show that the overlap part behaves as a superconductor $S'$ with an altered superconducting order parameter. In the last step, we review the appropriate matching conditions to apply at $x=0$ between the uncovered $S$-part and the fictitious $S'$-part, which are necessary due to the different transversal thicknesses.

\subsection*{First step: overlap geometry}
\begin{figure} 
\centering
  \def\svgwidth{1.08\columnwidth}
 \executeiffilenewer{InfOverlapp.svg}{InfOverlapp.pdf}%
 {inkscape -z -D --file=InfOverlapp.svg %
 --export-pdf=InfOverlapp.pdf --export-latex}%
 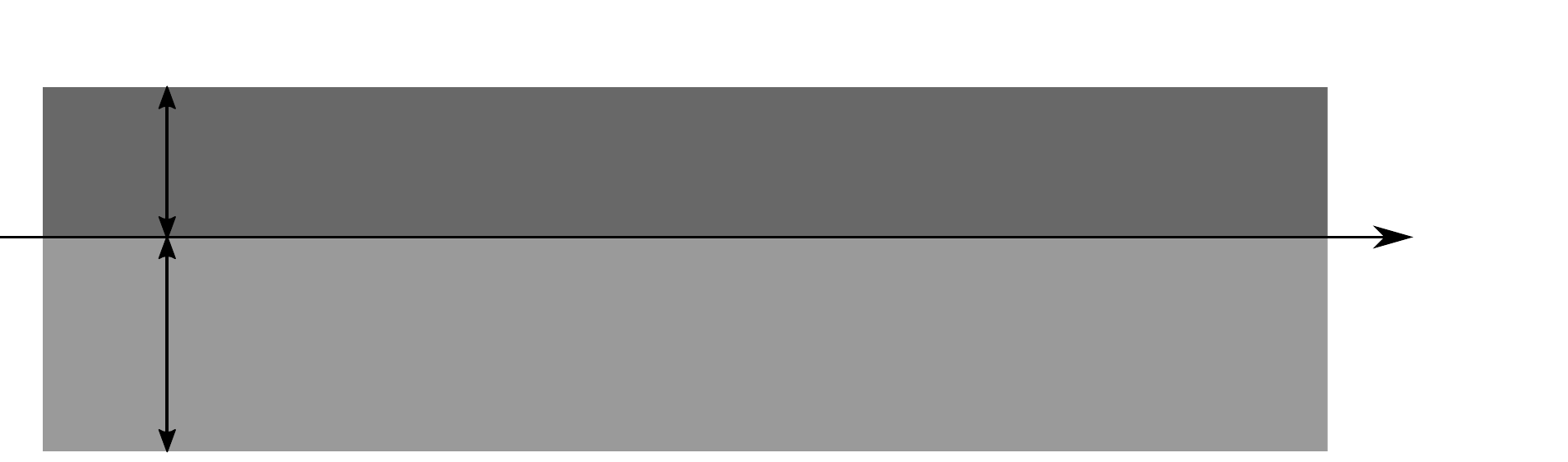
  \caption{Overlap geometry composed of two superconducting metals $S_{1/2}$ with thickness $d_{1/2}$ and superconducting order parameter $\Delta_{1/2}$.}
  \label{fig:InfOverlapp} 
\end{figure}
Consider an infinite overlap geometry composed of two superconducting metals $S_{1/2}$ with thickness $d_{1/2}$ and superconducting order parameter $\Delta_{1/2}$, as shown in \Fig~\ref{fig:InfOverlapp}. The Green's functions $\check{G}_{1/2}$ of the two metals each obey Usadel equations with associated $\Delta_{1/2}$ and boundary conditions: At $z=\pm d_{1/2}$ the normal derivative of $\check{G}_{1/2}$ vanishes, $\left. \partial_z \check{G}_{1/2}\right|_{z=\pm d_{1/2}}=0$, and at $z=0$ ones requires $\check{G}_1=\check{G}_2$ and $\partial_z\check{G}_1=\partial_z\check{G}_2$~\cite{kuprianov1988influence}, where $\partial_z$ denotes the partial derivative with respect to $z$. \\
For thicknesses small compared to the superconducting coherence lengths, $d_1,d_2\ll \xi_0=\text{min}\left\{ \xi_1,\xi_2 \right\}$ with $\xi_i = \sqrt{\hbar D/\Delta_i}$, the Green's functions can be expanded in a series in $z/\xi_0$. The boundary conditions are fulfilled by the expansion
\begin{align}
 \check{G}_{1/2}(x,z) \approx \check{G}_0(x) + \delta\check{G}(x)\left( \frac{z}{\xi_0} \pm \frac{z^2}{2d_{1/2} \xi_0} \right)
 \label{eq:Series}
\end{align}
with two yet unknown Green's functions $\check{G}_0,\delta \check{G}$. Note that the only $z$-dependence of $\check{G}$ comes from the last term, i.e. $\check{G}_0$ and $\delta \check{G}$ only depend on $x$. The Usadel equations for the two metals can now be used to find $\check{G}$ and $\delta \check{G}$: Plugging the above series \eq~\eqref{eq:Series} for either $\check{G}_1$ or $\check{G}_2$ into the associated Usadel equations, neglecting all terms with a $z$-dependence and assuming $\check{G}_0 - \left(d_{1(2)}/\xi_0\right)\delta \check{G} \approx \check{G}_0$ one finds for the correction
\begin{align*}
 \hbar D\delta\check{G}=\xi_0d_{1(2)}\check{G}_0\left\{ \hbar D\frac{\partial}{\partial x} \left( \check{G}_0 \frac{\partial \check{G}_0}{\partial x} \right) - \comm{-i E \check{\tau}_3 + \check{\Delta}_{1(2)}}{\check{G}_0} \right\},
\end{align*}
and using this expression in the other Usadel equation one obtains 1D Usadel equations for $\check{G}_0$
\begin{align}
\hbar D\frac{\partial}{\partial x} \left( \check{G}_0 \frac{\partial \check{G}_0}{\partial x} \right)=\comm{-i E\check{\tau}_3 +\check{\overline{\Delta}}}{\check{G}_0}
\end{align}
with $\check{\overline{\Delta}}=\left(d_1\check{\Delta}_1+d_2\check{\Delta}_2\right)/\left(d_1+d_2\right)$, which is just the over the transverse direction averaged superconducting order parameter. The retarded component gives the spectral Usadel equations already obtained in Refs.~\cite{belzig1999quasiclassical, fominov2001superconductive, hosseinkhani2018proximity}.

In the case of a normal metal there is no universal length scale such as the coherence length. Instead, one can introduce the energy dependent length scale $\xi_E = \sqrt{\hbar D/|E|}$. The above derivation then requires $d_N \ll \xi_E$ for all relevant energies. This gives the energy constraint $|E| \ll \sqrt{\hbar D/d_N^2} \equiv E_{\text{Th}}$ with the Thouless energy $E_{\text{Th}}$~\cite{thouless1977maximum}. The relevant energies for the spectral properties and transport processes in NIS-junctions at $T = 0$ are $\D$ and $e|V|$, which are both small compared to $\hbar \omega_D$ for the present study. For $d_N \ll \xi_0$ the associated $E_{\text{Th}}$ is comparable to $\hbar \omega_D$ and thus the above treatment of the overlap geometry is valid.

\subsection*{Second step: matching conditions}
Now, we return to the finite overlap geometry (see \fig~\ref{fig:setup}) and present matching conditions at $x=0$ for the Green's functions. How does the current density distribute transversely over the heterostructure? One might expect it to follow the path of least resistance by aggregating inside the superconductor. However, although a spatially non-continuous parameter (i.e. the superconducting order parameter) enters the exact two-dimensional Usadel equations, the distribution functions $f_{L/T}$ -- and thus also the current densities -- are, as solutions to a second-order differential equation, indeed continuous differentiable quantities. Consequently, in the small thickness limit, $d_1,d_2 \ll \xi_0$, with bordering hard walls, $\partial_z f=0$, the distribution functions have nearly no transverse dependence, and thus the current densities distribute homogeneously over the transverse direction as well. As current is conserved by the exact two-dimensional kinetic Usadel equations, one has to require for the approximate one-dimensional Usadel equations
\begin{align}
 d_1 \left.\frac{\partial f_{l}}{\partial x}\right|_{x=0}=(d_1+d_2)\left.\frac{\partial f_{r}}{\partial x}\right|_{x=0}, \label{eq:matchingCondKin}
\end{align}
in addition to the usual continuity $f_l=f_r$ at $x=0$. The subscripts $l/r$ refer to ``left'' and ``right'' with respect to the contact at $x=0$, where the two metals with different thickness meet.

There is no conserved quantity for the spectral Usadel equations in general. \footnote{In a superconductor the spectral supercurrent density $j_E=\Tr{\hat{\tau}_3\hat{R}\nabla\hat{R}}$ (which is not be confused with the energy current density also denoted by $j_E$) is not conserved in general.} In order to reduce the number of variables in a system, one usually averages the Green's functions $\check{G}(\textbf{r})$ over the silent directions the Green's functions do not depend on. Consequently, at locations where metals with different cross-sectional area are in contact, the so defined new Green's functions are not continuous differentiable. To see that, consider e.g. the present setup \fig~\ref{fig:setup}:
\begin{align}
 &\left. \frac{\partial \check{G}_l}{\partial x}\right|_{x=0} = \frac{1}{d_1}\int \limits_{0}^{d_1} \left.\frac{\partial \check{G}}{\partial x}\right|_{x=0} \mathrm{d}z \\
 &\left. \frac{\partial \check{G}_r}{\partial x}\right|_{x=0} = \frac{1}{d_1+d_2}\left[\int \limits_{0}^{d_1} \left.\frac{\partial \check{G}}{\partial x}\right|_{x=0}\mathrm{d}z + \int \limits_{d_1}^{d_2} \underbrace{\left.\frac{\partial \check{G}}{\partial x}\right|_{x=0}}_{=\,0} \mathrm{d}z\right] \\
 \Rightarrow& d_1 \left. \frac{\partial \check{G}_l}{\partial x}\right|_{x=0} = \left(d_1+d_2\right) \left. \frac{\partial \check{G}_r}{\partial x}\right|_{x=0} \label{eq:matchingCondGen},
\end{align}
where the last integral in the second line vanishes due to the hard-wall boundary condition. 

Together with the usual continuity $\check{G}_l = \check{G}_r$ at $x=0$ the matching conditions \eq~\eqref{eq:matchingCondGen} describe the local conservation law of the spectral current $\check{G}\partial_x\check{G}$. They can be generalized taking into account arbitrary cross-sectional areas and conductivities for the two metals in contact~\cite{kuprianov1988influence}. Furthermore, the Keldysh components of \eq~\eqref{eq:matchingCondGen} imply the matching conditions \eq~\eqref{eq:matchingCondKin} for the distribution functions.

\subsection*{Parameterization of Green's functions}
The normalization requirement $\check{G}\check{G}=\check{1}$ for the Keldysh Green's functions reads in terms of $\hat{R},\hat{A}, \hat{K}$
\begin{center}
\begin{align*}
 \hat{R}\hat{R} = \hat{A}\hat{A} = \hat{1} \\
 \hat{R}\hat{K} + \hat{K}\hat{A} = 0.
\end{align*}
\end{center}
It thus allows for the so called trigonometric $\theta$-parameterization
\begin{align*}
 \hat{R} &= \begin{pmatrix} \cos{\theta} & \sin{\theta} \textrm{e}^{i \phi} \\ \sin{\theta} \textrm{e}^{-i \phi} & - \cos{\theta} \end{pmatrix} \\
 \hat{A} &= - \hat{\tau}_3 \hat{R}^{\dagger} \hat{\tau_3},~~~~~
 \hat{K} = \hat{R}\hat{h} - \hat{h}\hat{A}
\end{align*}
with two complex quantities $\theta(E, x)$ and $\phi(E, x)$. The matrix $\hat{h}$ is related to the distribution functions for electrons $f_e$ and holes $f_h$ by
\begin{align*}
 \hat{h} = \begin{pmatrix} 1 - 2f_e & 0 \\ 0 & 2 f_h - 1 \end{pmatrix}.
\end{align*}
It is convenient to split the distribution matrix into odd and even component with respect to the Fermi surface, $\hat{h} = f_L + f_T \hat{\tau}_3$, where $f_{L/T}$ refer to the ``longitudinal'' and ``transverse'' modes, respectively~\cite{schmid1975linearized, tinkham2004introduction}.

In units of the coherence length $\xi_0 = \sqrt{\hbar D/\Delta_{\text{BCS}}}$ and $\Delta_{\text{BCS}}$ the Usadel equations~\eqref{eq:Usadel_Keld} and self-consistency equation~\eqref{eq:Delta_Keld} read
\begin{align}
 \frac{1}{2} \frac{\partial^2\theta}{\partial x^2} + \left[ i \e - \frac{1}{2} \cos{\theta} \left( \frac{\partial \phi}{\partial x} \right)^2 \right]\sin{\theta} + \\ \cos{\theta} \frac{\Delta \textrm{e}^{-i \phi} + \Delta^* \textrm{e}^{i \phi}}{2} &= 0 \label{eq:spec_Usadel1}\\
 \frac{\partial}{\partial x} \left( -i \text{sin}^2\left( \theta \right) \frac{\partial \phi}{\partial x}  \right) - \sin{\theta}\left( \Delta \textrm{e}^{-i \phi} - \Delta^* \textrm{e}^{i \phi} \right) &= 0 \label{eq:spec_Usadel2}
\end{align}

\begin{align}
 &\frac{\partial j_E}{\partial x} = 0, ~~~~~~~~~~~~~~~~~~~~~~~~~ j_E=\mathcal{D}_L \frac{\partial f_L}{\partial x} - \mathcal{T}\frac{\partial f_T}{\partial x} + j_S f_T \label{eq:kinetic1}\\
 &\frac{\partial j_C}{\partial x} = \mathcal{R} f_T - \mathcal{L} f_L, ~~~~j_C=\mathcal{D}_T \frac{\partial f_T}{\partial x} + \mathcal{T}\frac{\partial f_L}{\partial x} + j_S f_L \label{eq:kinetic2}
\end{align}

\begin{align}
 \Delta = \frac{\lambda}{2 i} \int \limits_{0}^{\frac{\hbar \omega_D}{\D}} \left[\sin{\theta}\textrm{e}^{i \phi}(f_L - f_T)-\sin{\theta^*}\textrm{e}^{i\phi^*}(f_L+f_T)\right]\mathrm{d}\e. ~\label{eq:selfconsUsadel}
\end{align}

The energy dependent coefficients appearing in the kinetic equations~\eqref{eq:kinetic1}-\eqref{eq:kinetic2} are given by
\begin{align*}
 \mathcal{D}_L &= \frac{1}{4} \tr{\hat{\tau}_0 - \hat{R}\hat{A}} \\ &= \frac{1}{2}\left[ 1 + \abs{\cos{\theta}}^2 - \abs{\sin{\theta}}^2 \cosh{2 \Im{\phi}} \right]\\
  \mathcal{D}_T &= \frac{1}{4} \tr{\hat{\tau}_0 - \hat{R}\hat{\tau}_3\hat{A}\hat{\tau}_3} \\ &= \frac{1}{2}\left[ 1 + \abs{\cos{\theta}}^2 + \abs{\sin{\theta}}^2 \text{cosh}\left( 2 \text{Im}\left\{ \phi \right\} \right) \right] \\
  j_S &= \frac{1}{4} \tr{\hat{\tau}_3\left(\hat{R}\frac{\partial \hat{R}}{\partial x} - \hat{A}\frac{\partial \hat{A}}{\partial x} \right)} = \text{Im}\left\{ \text{sin}^2\!\left( \theta \right)\,\frac{\partial \phi}{\partial x} \right\} \\
  \mathcal{T} &= \frac{1}{4}\tr{\hat{\tau}_3\hat{A}\hat{R}} = \frac{1}{2}\left|\sin{\theta}\right|^2 \text{sinh}\left(2\text{Im}\!\left\{ \phi \right\}\,\right) \\
  \mathcal{L} &= \frac{1}{2}\tr{\hat{\tau}_3 \hat{\Delta}\left( \hat{A} - \hat{R} \right)} = -\text{Re}\left\{ \sin{\theta} \left(\Delta \textrm{e}^{-i\phi} - \cc{\Delta} \textrm{e}^{i \phi} \right) \right\} \\
  \mathcal{R} &= \frac{1}{2}\tr{ \hat{\Delta}\left( \hat{R} + \hat{A} \right)} = \text{Re}\left\{ \sin{\theta} \left(\Delta \textrm{e}^{-i\phi} + \cc{\Delta} \textrm{e}^{i \phi} \right) \right\}
\end{align*}
Here, $\mathcal{D}_L$ and $\mathcal{D}_T$ play the role of normalized diffusion coefficients for the energy ($j_E$) and charge ($j_C$) current densities, respectively, $\mathcal{T}$ is a cross-diffusion term and $j_S$ gives the density of supercurrent carrying states. The coefficients $\mathcal{R}$ and $\mathcal{L}$ are related to a leakage current, whereas $\mathcal{R}$ in particular is connected to Andreev reflection~\cite{andreev1964thermal}.

These equations must be supplemented with appropriate boundary conditions. The Kuprianov-Lukichev boundary conditions~\cite{kuprianov1988influence} at the tunnel junction, $x=-L_1$, read using the above parameterization
\begin{align}
 r \frac{\partial \theta}{\partial x} &= \sin{\theta} \\
 \frac{\partial \phi}{\partial x} &= 0 \\
 r \mathcal{D}_{L/T} \frac{\partial f_{L/T}}{\partial x} &= N_S\left[ f_{L/T} - f_{L/T}^{(R)} \right],
\end{align}
with the distribution functions of the reservoir
\begin{align*}
 f_{L/T}^{(R)} = \frac{1}{2}\left[ \text{sign}(E+eV)\pm \text{sign}(E-eV) \right].
\end{align*}

At the grounding, $x = L_2$, the distribution functions recover their equilibrium value, $f_L = \text{sign}(E), f_T = 0$, whereas hard-wall boundary conditions $\partial_x \hat{R} = 0$ are assumed for the spectral quantities, which read $\partial_x\theta = \partial_x \phi = 0$.

From the kinetic equations~\eqref{eq:kinetic1}-\eqref{eq:kinetic2}, it is evident that the spectral energy current, and thus also the physical energy current $J_E \propto \int E j_E \mathrm{d}E$ is conserved. The spectral charge current, however, is not conserved. Instead, the leakage current $j_{\text{leak}} = \mathcal{R}f_T - \mathcal{L}f_L$ describes its spectral redistribution (see \fig~\ref{fig:CurrentConversion}). The conservation of the physical charge current is not so obvious: With the explicit definitions of $\mathcal{R}$ and $\mathcal{L}$ and the order parameter \eq~\eqref{eq:selfconsUsadel}, the energy integral of the RHS of \eq~\eqref{eq:kinetic2} can be rewritten as

\begin{align*}
 &\int \limits_{0}^{\frac{\hbar \omega_D}{\D}} (\mathcal{R}f_T - \mathcal{L}f_L) \d{\e} = \\ &\int\limits_{0}^{\frac{\hbar \omega_D}{\D}}\Re{\de \sin{\theta}\exp{i\phi}(f_L+f_T)-\cc{\de}\sin{\theta}\exp{i\cc{\phi}}(f_L-f_T)}\d{\e}\\ = &\,2\,\Re{- i\frac{\abs{\de}^2}{\lambda}+i\int \limits_{0}^{\frac{\hbar \omega_D}{\D}} \Im{\de \sin{\theta}\exp{-i\phi}}(f_L+f_T)\d{\e}} = 0.
\end{align*}
Hence, the physical charge current is indeed conserved, $\partial_x J_C \propto \int \partial_x j_C \d{\e} = \int j_{\text{leak}}\d{\e} = 0$.

\section{Numerical results}
\label{sec:NumRes}
If not mentioned otherwise we take as parameters $L_1 = L_2 = 2 \xi_0$ for both setups, and $d_N = d_S \ll \xi_0$ for \SWT.

\subsection*{Order parameter}

\begin{figure}
 \centering
 \includegraphics[width=\columnwidth, clip, keepaspectratio]{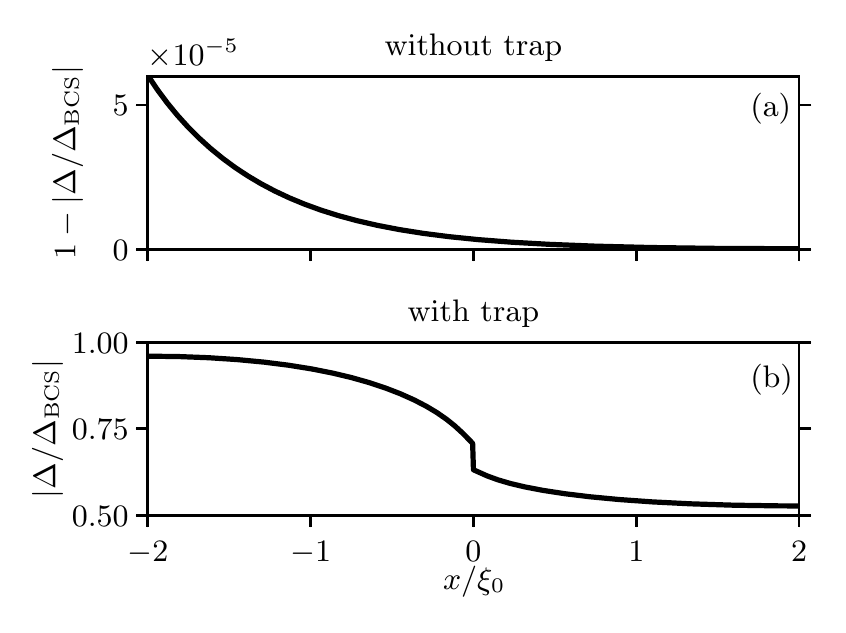}
 \caption{Spatial profile of the superconducting order parameter along the superconductor, computed for $eV = 0$. For \SWOT~(top panel (a)) the inverse proximity effect of the normal-metal reservoir on the superconductor is almost negligible due to the tunnel barrier, hence the order parameter only slightly deviates from the bulk value. The good metallic contact of the normal metal in \SWT~(bottom panel (b)) leads to a halving of the order parameter (for $d_S = d_N$) of the underlying superconducting part, which also influences the rest of the superconductor. Note that at the injector with a distance of $L_1 = 2$ to the normal-metal trap, the superconducting order parameter recovers to over 95 \% of its bulk value.}
 \label{fig:OPs}
\end{figure}
For \SWOT, the inverse proximity effect from the normal-metal reservoir only leads to a slight spatial modification of the superconducting order parameter, as seen in \fig~\ref{fig:OPs} (a). 

\Fig~\ref{fig:OPs} (b) shows the great impact of the proximity effect of the normal-metal trap on the superconductor: It reduces the order parameter of the underlying superconductor to a bulk value of $d_S\D/(d_S + d_N)$, which, in turn, leads to a reduction of the order parameter of the uncovered superconducting part. Their mutual adjustment leads to a stronger bending on the uncovered site, since the SN bilayer has a greater thickness of $d_S + d_N$. Moreover, the discontinuity at $x=0$ in the superconducting order parameter and the electron pairing interaction strength $\lambda$ is such that the pairing amplitude $F = \lambda \Delta$ is continuous.

Furthermore, for both setups a dependence on the applied voltage, due to QP injection or formation of a supercurrent, is hardly discernible, which is in accordance with numerical results showing that $n_{\text{qp}}/(N_0 \D) \ll 1$ and $J_S/J_{\text{crit}} \ll 1$ with the critical supercurrent density $J_{\text{crit}} \approx \frac{3}{4} \D \sigma_N/(e \xi_0)$ \cite{anthore2003density} (see \fig~\ref{fig:CompareN_qp(V)} and \fig~\ref{fig:CompareJ_C} below).

\subsection*{Density of states}
 \begin{figure}
  \begin{center}
  \includegraphics[width=\columnwidth, clip, keepaspectratio]{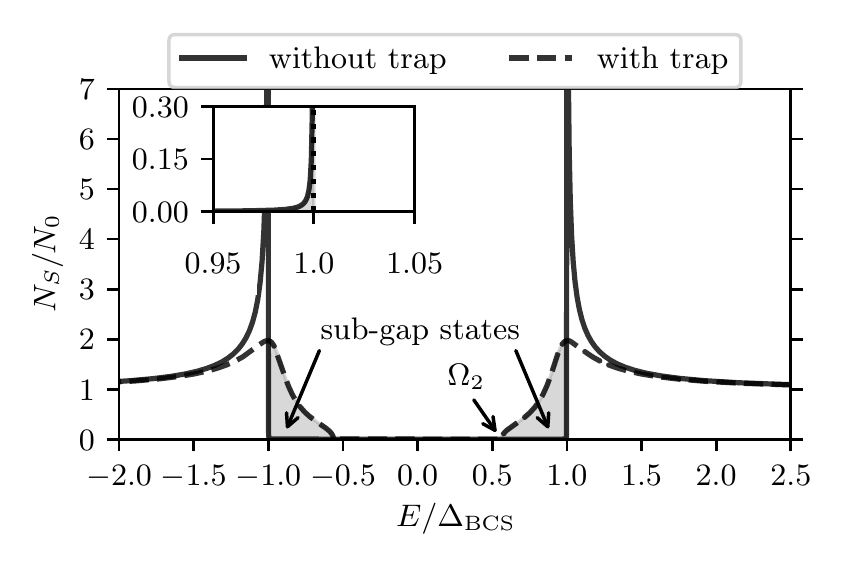}
  \caption{Comparison of the superconducting DOS at the QP injector ($x=-L_1$) for \SWT~(dashed line) and \SWOT~(solid line), computed for $eV = 0$. The respective deviations from the BCS bulk DOS (dotted line, almost entirely covered by solid line) result from the proximity effect. There, the influence of the normal-metal reservoir is again almost negligible due to the tunnel barrier (see inset). For \SWT, the proximity effect from the normal-metal trap results in a significant reduction of the spectral energy gap to a value $\Omega_2 \approx 0.56\D$, accompanied by a reduction of the BCS peak at an energy $|E| = \D$, while the superconducting order parameter $\Delta$ almost recovers its BCS bulk value (see \fig~\ref{fig:OPs} (b)). The sub-gap states, depicted by the grey-shaded areas, with energies $|E| < \D$, for which the bulk DOS vanishes, are additionally available for occupation.}
  \label{fig:CompareDOS}
  \end{center}
 \end{figure}
Within the Usadel formalism, the QP DOS can be computed from the retarded Green's function via
\begin{align*}
 N_S(E, x) = \frac{N_0}{2} \Re{\Tr{\hat{R}(E, x) \hat{\tau}_3}} = N_0 \Re{\cos{\theta(E, x)}}.
\end{align*}

\Fig~\ref{fig:CompareDOS} shows the superconducting DOS for both setups at the QP injector. For \SWOT~(dashed line) the DOS almost coincides with that of a BCS bulk superconductor (dotted line). The inverse proximity effect from the normal-metal reservoir on the superconductor is strongly suppressed due to the tunnel barrier and only leads to a slight broadening of the BCS energy gap with a small but non-vanishing DOS for energies $|E| \leq \Delta_{\text{BCS}}$, revealing the existence of sub-gap states, i.e. states with energy $|E| < \D$ for which the BCS DOS vanishes (see inset of \fig~\ref{fig:CompareDOS}).

\begin{figure}
 \centering
 \includegraphics[width=\columnwidth, clip, keepaspectratio]{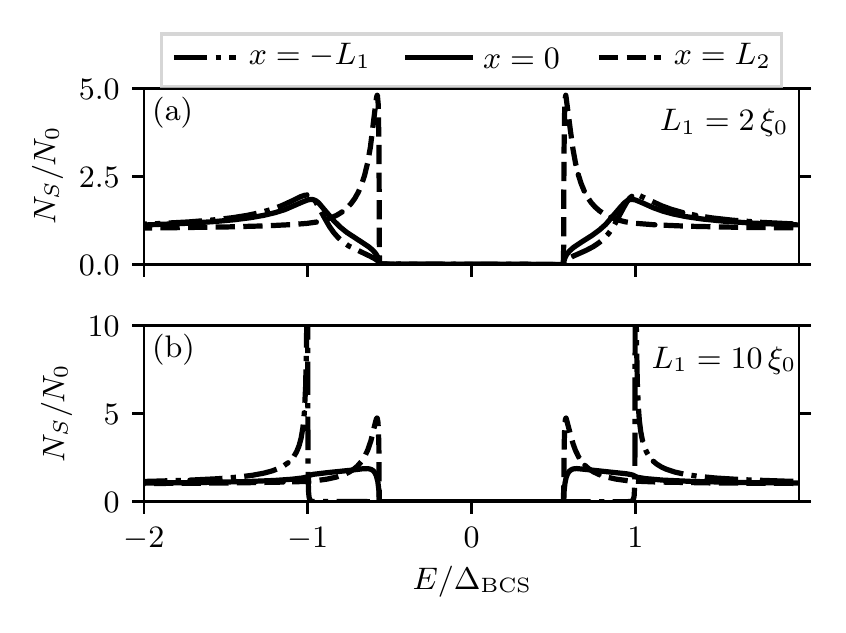}
 \caption{Superconducting DOS at the QP injector (dashdotted line), side of the trap facing the injector (solid) and electrical grounding (dashed line) for a distance of $L_1 = 2 \xi_0$ (top panel (a)) and $L_1 = 10 \xi_0$ (bottom panel (b)) between the QP injector and the normal-metal trap (see \fig~\ref{fig:setup} for setup details), computed for $eV = 0$. The DOS approaches the BCS bulk behavior with increasing distance from the trap, hence losing the striking features of \SWT~with $L_1=2\xi_0$ (see \fig~\ref{fig:CompareDOS}), which are thus caused by the proximity effect from the normal-metal trap.}
 \label{fig:CompareDOSShortLong}
\end{figure}

For \SWT~the pronounced reduction of the spectral energy gap $\Omega_2$ in the DOS with a significant increase in the number of sub-gap states and the reduction of the peak at $|E|=\Delta_{\text{BCS}}$ are most salient. As \fig~\ref{fig:CompareDOSShortLong} illustrates, this is traced back to the close proximity of the normal-metal trap ($L_1=2\xi_0$ for top panel (a)) with a distance of $L_1 = 2\xi_0$, as these features almost recover their BCS bulk behavior for $L_1=10\xi_0$ (see bottom panel (b)). Note that this in contrast to the superconducting order parameter, which almost recovers its BCS bulk value at the injector for $L_1 = 2\xi_0$ (see \fig~\ref{fig:OPs}). Such discrepancy between the spectral energy gap $\Omega$ in the DOS and the absolute value of the order parameter $|\Delta|$ are known from gapless superconductivity, which can occur in both equilibrium and non-equilibrium situations~\footnote{See, for example~\cite{tinkham2004introduction, de2018superconductivity, phillips1963gapless} and references therein.} and in hybrid structures in thermal equilibrium with striking agreement between experiment and theory based on the Usadel formalism~\cite{cherkez2014proximity}.

See \fig~\ref{fig:CompareDOSShortLong} (a) for the DOS at different positions within the superconductor.

\subsection*{Quasiparticle injection}
The density of populated QP states $\n = n_h + n_e$ has contributions
\begin{align*}
 n_h(x) = \int \limits_{-\hbar \omega_D}^{0} N_S(E, x) f_h(E, x) \d E
\end{align*}
from hole-like excitations with $E < 0$ and 
\begin{align*}
 n_e(x) = \int \limits_{0}^{\hbar \omega_D} N_S(E, x) f_e(E, x) \d E
\end{align*}
from electron-like excitations with $E > 0$. Using the particle-hole symmetry $N_S(-E) = N_S(E)$ and $f_h(-E) = f_e(E)$, the total density of QPs can be written as
\begin{align}
 \n(x) = \int \limits_{0}^{\hbar \omega_D} N_S(E, x) \left[ 1 - f_L(E, x) - f_T(E, x) \right]\d E. \label{eq:n_qp}
\end{align}
At the grounding, the distribution functions recover their zero-temperature equilibrium values, $f_L(E) = \text{sign}(E), f_T(E) = 0$, so that the QPs are forced to vanish there, $\n(x=L_2) = 0$.
\begin{figure}
 \centering
 \includegraphics[width=\columnwidth, clip, keepaspectratio]{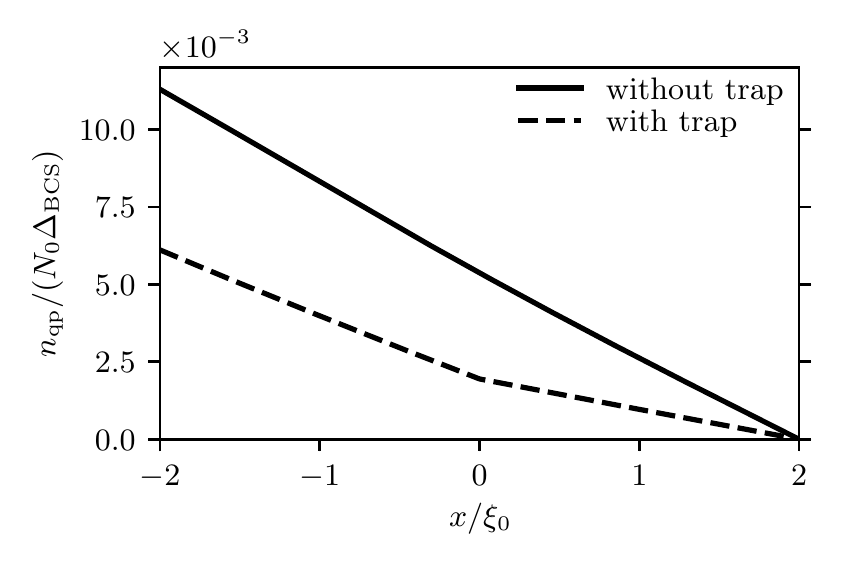}
 \caption{Comparison of the QP density along the superconductor with (solid line) and without (dashed line) trap for a voltage $e|V| = 10 \Delta_{\text{BCS}}$. With attached trap the QP density is reduced throughout the superconductor. For \SWOT, the superconductor is almost homogeneous due to the negligible proximity effect, and therefore the diffusion of the QPs through the superconductor result in a linear curve, in accordance with the approximate solutions given in the appendix. The kink in the dashed line for \SWT~results from the changing thicknesses at $x=0$ (see \fig~\ref{fig:setup}), entering via the effective matching conditions \eq~\eqref{eq:matchingCondGen}.}
 \label{fig:CompareN_qp(x)}
\end{figure}
\Fig~\ref{fig:CompareN_qp(x)} shows the spatial profile of the QP density $\n(x)$ along the superconductor for an applied voltage of $e|V|=10\D$. The trap leads to a reduction of the QP density throughout the superconductor. For \SWOT~the proximity effect is strongly suppressed. Consequently, the superconductor is almost homogeneous, the spectral coefficients that enter the kinetic Usadel equations~\eqref{eq:kinetic1}-\eqref{eq:kinetic2} are spatially independent, and the diffusion of the QPs through the superconductor leads to a linear change in the QP density. The approximate solutions, which assume a homogeneous superconductor and are given in the appendix, are in good agreement with the numerical results.

The presence of the sub-gap states in the DOS \fig~\ref{fig:CompareDOS} makes a QP injection possible for voltages $e|V| < \D$ (see \fig~\ref{fig:CompareN_qp(V)}). 

\subsection*{Trapping performance}
\begin{figure}
 \centering
 \includegraphics[width=\columnwidth, clip, keepaspectratio]{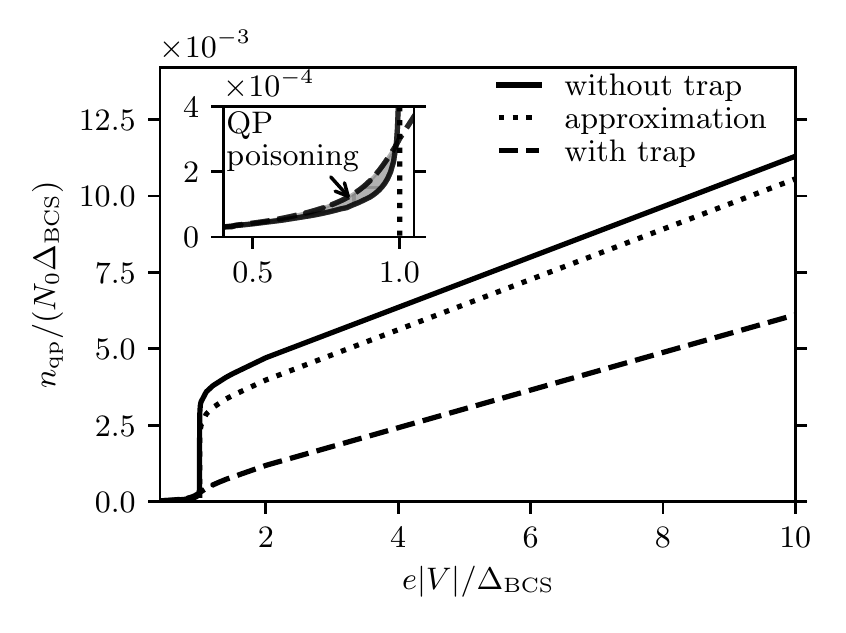}
 \caption{QP density at the injector as function of applied voltage with (dashed line) and without (solid line) trap. The two striking features of the DOS for \SWT~(see \fig~\ref{fig:CompareDOS}), caused by the proximity effect, influence the efficiency of the trap competitively: While the reduction of the DOS at an energy $|E| = \D$ leads to the strongly reduced QP density at an injection voltage $e|V| = \D$, the new available sub-gap states lead to QP poisoning, i.e. their occupation at lower voltages. The off-set between the parallel linear curves at higher voltages is explained below by the conversion between dissipative normal and supercurrent.}
 \label{fig:CompareN_qp(V)}
\end{figure}

The trapping performance can be demonstrated and quantified by a direct comparison of the density of injected QPs for both setups, see \fig~\ref{fig:CompareN_qp(V)}: At a voltage $e|V|$ slightly above $\D$, the QP density for \SWOT~is bigger  than that for \SWT~by a factor of approximately 7.6. This is due to the inverse proximity effect, which leads to a significant reduction in the DOS at $|E| = \D$ \footnote{The QP density is not only controlled by the DOS, but also by the QP distribution function. The distribution function shows a step-like behaviour, where the width of the middle-step coincides with the according gap in the DOS, so that the departure from the equilibrium distribution does not manifest itself. The agreement of the numerical results for the the QP and current density with that given in \cite{tinkham2004introduction}, which assume an equilibrium distribution, support this finding.} (see \fig~\ref{fig:CompareDOS}). However, since the total number of available states is not altered by the (inverse) proximity effect, 
\begin{align*}
 \int \limits_{-\hbar \omega_D}^{\hbar \omega_D} \left( N_S - N_0 \right) \d{E} = 0,
\end{align*}
the reduction of the $|E|=\D$ peak is accompanied by a softening of the spectral energy gap down to $\Omega \approx 0.56 \D$ with the existence of the sub-gap states. This feature is pronounced much more significantly for \SWT, which leads to QP poisoning for voltages $e|V| < \D$, i.e. higher QP densities. Note, however, that the ratio of QP densities for \SWT~and \SWOT~is not higher than approximately 1.4, even though the DOS differ significantly from each other for energies $|E| < \D$.

The location of the trap plays a decisive role in the trapping performance as the superconducting DOS recovers its bulk-form with increasing distance to the trap. Consequently, in the limit $L_1 \gg \xi_0$ the trap does not have an impact on the injection and density of the QPs. The opposite limit $L_1=0$ is equivalent to \SWOT~but with the superconductor $S$ replaced by $S'$ with half the initial spectral energy gap. The resulting injection curve $\n(V)$ at the injector is obtained from the solid line in \fig~\ref{fig:CompareN_qp(V)} horizontally shifted by $0.5$ units, indicating a QP poisoning for all voltages. These two limits clearly show the existence of a trap position with optimal trapping performance.

The integrand in \eq~\eqref{eq:n_qp} is almost independent of the applied voltage (apart from the fact that $f(|E|>e|V|)=0$), and almost equal for both setups at high energies $|E| \gg \D$. This, together with the finding that the integrand is strongly peaked at $|E| = \D$ due to the DOS for \SWOT, explains why both curves are almost parallel with an offset of approximately $5.2 \times 10^{-3} N_0 \D$ for voltages $e|V| \gtrsim 2\D$. This offset depends on the setup geometries and tends to zero in the limit $L_1 \gg \xi_0$. Below, we will qualitatively explain the appearance of this offset via the conversion between normal and supercurrent.

\subsection*{Current conversion}
The Usadel formalism allows for a spectral resolution of the physical charge and energy current densities, $J$, in terms of their respective spectral ones, $j$:

\begin{align}
J_C &= \frac{\sigma_N}{e} \int \limits_{0}^{\hbar \omega_D} j_C \d E\\
 J_E &= -\frac{\sigma_N}{e^2} \int \limits_{0}^{\hbar \omega_D} E j_E \d E
\end{align}

with the normal-state conductivity $\sigma_N = e^2 N_0D$ of the metal and the current densities given in \eq~\eqref{eq:kinetic1}-\eqref{eq:kinetic2}. The dissipative part is proportional to the gradient of the distribution functions, whereas the last term accounts for the supercurrent, respectively.

\begin{figure}
 \centering
 \includegraphics[width=\columnwidth, clip, keepaspectratio]{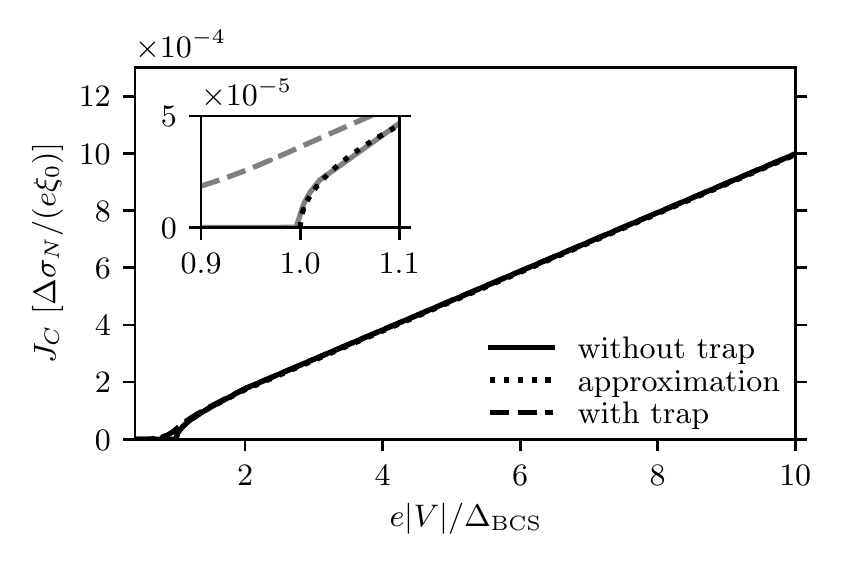}
 \caption{Current-voltage characteristics for the setup with (dashed line) and without (solid line) normal-metal trap. The dotted line in the inset shows the theoretically predicted $I-V$ curve in the diffusive limit with neglected proximity effect. The deviations from this curve for the present setups are due to the sub-gap states, which are occupied and contribute to the charge current for voltages $e|V|<\D$.}
 \label{fig:CompareJ_C}
\end{figure}

The current-voltage characteristics of both setups are shown in \fig~\ref{fig:CompareJ_C}.

\begin{figure}
 \centering
 \includegraphics[width=\columnwidth, clip, keepaspectratio]{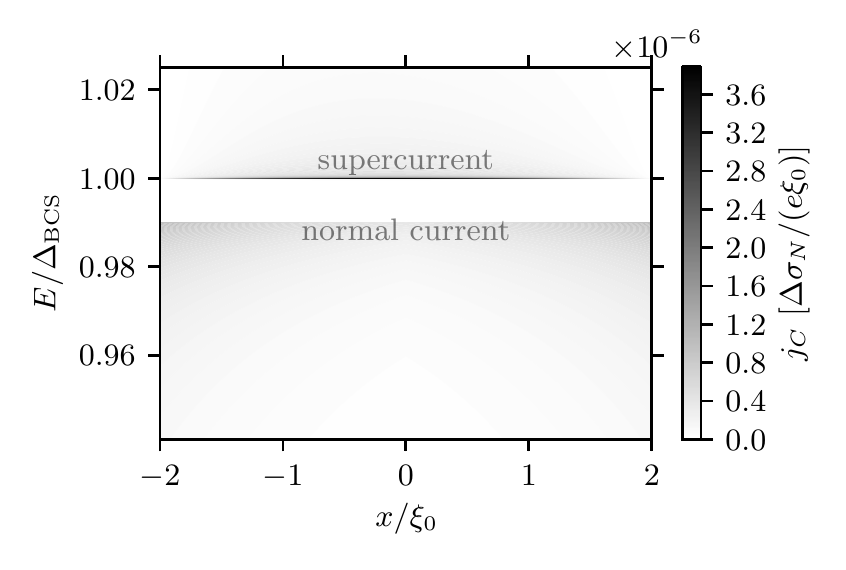}
 \caption{Spectrally resolved charge current with contributions to the dissipative normal current and supercurrent as a function of position (horizontal axis) and energy (vertical axis) at a voltage $e|V| = 0.99 \D$ just below the spectral energy gap for \SWOT.}
 \label{fig:specChargeCurr}
 \end{figure}

\Fig~\ref{fig:specChargeCurr} shows a contour plot of the spectral charge current $j_C$ along the superconductor without normal-metal trap in the relevant energy interval for an applied voltage of $e|V| = 0.99\D$. The sub-gap states present in the DOS \fig~\ref{fig:CompareDOS} make a QP injection and a current flow possible for voltages $e|V| < \D$. The energies of QPs entering the superconductor and thus the spectral contributions to the normal current are bounded by $e|V|$, whereas the states with energies $|E| \gtrsim \D$ contribute to the supercurrent, most significantly at the peak of the DOS (see \fig~\ref{fig:CompareDOS}) and regardless of the applied voltage. Note that these two contributions overlap for voltages $e|V| \gtrsim \D$.

\begin{figure}
 \centering
 \includegraphics[trim=53mm 105mm 70mm 135mm, width=\columnwidth, clip, keepaspectratio]{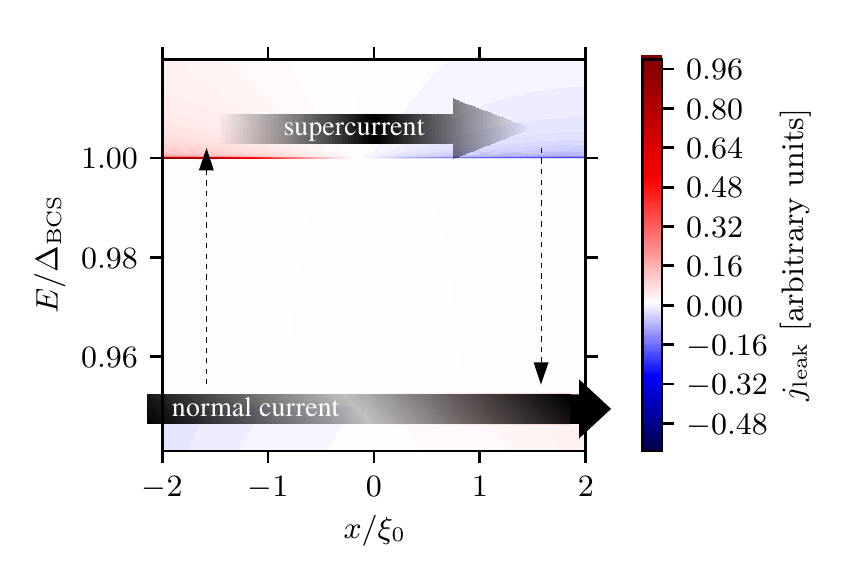}
 \caption{Contour plot of the leakage current without normal-metal trap at a voltage $e|V| = 0.95 \D$. Shown is the spectral redistribution of the charge current density (see \fig~\ref{fig:specChargeCurr}) in the relevant energy interval and the partial conversion between normal and supercurrent along the superconductor. The transparency level of the arrows indicate the amount of the respective current to the total charge current (see \fig~\ref{fig:specChargeCurr}).}
 \label{fig:CurrentConversion}
\end{figure}

From \eq~\eqref{eq:kinetic2} it is evident, that the spectral charge current is not conserved in a superconductor. Instead, the leakage current leads to its spectral redistribution. This process is visualized in \fig~\ref{fig:CurrentConversion}: According to \fig~\ref{fig:specChargeCurr}, the charge current entering the superconductor at the injector is entirely made out of dissipative normal current. While passing through the superconductor, the spectral charge current gets shared among states with energies $|E| \leq e|V|$ and $|E| \approx \D$ indicated by the blue ($j_{\text{leak}} < 0$) and red ($j_{\text{leak}} > 0$) areas.  This manifests itself in an increase of the supercurrent and a decrease of the normal current (see also \fig~\ref{fig:specChargeCurr}), respectively, indicated by the varying transparency of the associated arrows. This conversion happens on a length scale of about $2 ~\xi_0$, after which the whole process is almost reversed. \footnote{Note the lack of symmetry around $x=0$, which is due to the unsymmetrical boundary conditions.} Note, however, that the current conversion takes place in a normal metal as well, which therefor cannot be determined solely by the leakage current, since it vanishes in a normal metal due to $\de = 0$.

The purely normal charge current entering the superconductor,
\begin{align*}
 J_C=\frac{\sigma_N }{e} \int \limits_{0}^{e|V|} \left.\left( \mathcal{D}_T \frac{\partial f_T}{\partial x} + \mathcal{T} \frac{\partial f_L}{\partial x} \right)\right|_{x=-L_1}\d{E},
\end{align*}
is carried by states with an energy up to $e|V|$. The lower boundary in the above integral must be effectively replaced by the spectral energy gap in the DOS, as (almost) no states are available for occupation below it.

\begin{figure}
 \centering
 \includegraphics[width=\columnwidth, clip, keepaspectratio]{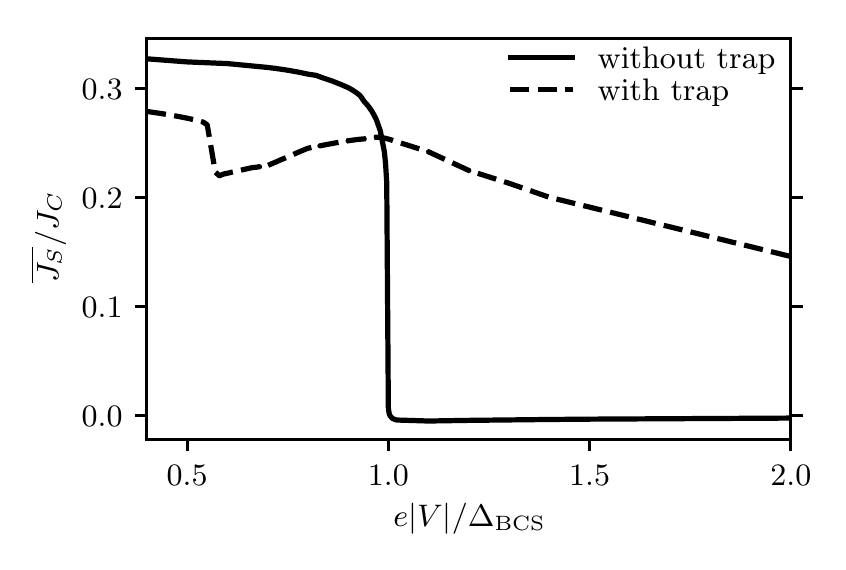}
 \caption{Supercurrent density averaged along the superconductor and normalized to the total charge current density as a function of applied voltage for \SWT~(dashed line) and \SWOT~(solid line). The conversion between dissipative normal current and supercurrent is due to Andreev reflection of states with energy $|E| \leq |\de|$, which takes place over a length $\sim 2\xi_0$. This, and the amount of sub-gap states, i.e. the discrepancy of the spectral energy gap $\Omega$ in the DOS and $|\de|$, explain the different ratios of conversion into supercurrent.}
 \label{fig:MaxConversion}
\end{figure}

The conversion between normal and supercurrent is due to Andreev reflection~\cite{andreev1964thermal} of states with energy $|E|<|\de|$ described by $\mathcal{R}$. Note that the spectral energy gap $\Omega$ in the DOS might differ from $\de$, as it is the case for \SWT. For \SWOT~with a negligible proximity effect, $\mathcal{R}$ almost attains its BCS bulk value and thus vanishes for energies $|E| \gtrsim \D$. But even for \SWT~with a non-negligible impact of the proximity effect, the order parameter has a magnitude close to unity at the injector and decreases monotonically throughout the superconductor (see \fig~\ref{fig:OPs} (a)). The spectral energy gap $\Omega$ is significantly reduced and the DOS is clearly non-vanishing for states with energy $|E| > \Omega$ due to the proximity effect, so that these new states contribute to the charge current. Since $\mathcal{R}$ is non-vanishing for these energies, the associated states get Andreev reflected and thus contribute to the supercurrent. This explains why the conversion from normal to supercurrent is so pour for \SWOT~compared to \SWT~for voltages $e|V| > \D$ (see \fig~\ref{fig:MaxConversion}). For voltages $e|V| < \D$ all occupied states get Andreev reflected and thus contribute to supercurrent, giving rise to the sudden jump at $e|V| = \D$. This conversion process is not local, but instead takes place over a length of about $2\xi_0$. In addition, the grounding at $x=L_2$ forces an entire reconversion from super- to normal current, so that both setups with a total length $L_S=L_1+L_2 = 4\xi_0$ of the superconductor each are too short for a pronounced conversion close to unity. This might also explain why the conversion for \SWOT~is higher than for \SWT~for voltages $e|V| < \D$. This could be resolved by increasing the length of the superconducting part, leading to a decline in conversion for voltages $e|V|>\D$.
\subsection*{Andreev reflection and QP reduction}
The mutual conversion between normal and supercurrent via Andreev reflection~\cite{jacobs2001dynamics} affects the QP density: The dissipative \normalcurrent~is due to a diffusive motion of the QPs and is thus almost proportional to the gradient of their density, $J_N \propto \nabla \n$. Consequently, the more normal current is converted into supercurrent along the superconductor, the more the QP density gradient decreases. A pronounced conversion, combined with the electrical grounding draining the QPs $\n = 0$, leads to a reduction of the QP density throughout the whole superconductor. 

This rather qualitative view can be made more quantitative: Integrating $\partial_x \n = \alpha J_N = \alpha(J_C - J_S)$ with a phenomenological proportionality factor $\alpha$ along the superconductor yields for the QP density at the injector $\n(x=-L_1) = \alpha L\left( J_C - \overline{J_S} \right)$, where $\overline{J_S}$ denotes the supercurrent averaged along the superconductor and it was used that $n(x=L_2)=0$ due to the electrical grounding. From the numerical solutions for high voltages $e|V| \gg \D$ the factor $\alpha$ is found to be approximately $\alpha^{(1)}\approx2.83$ for \SWOT~and $\alpha^{(2)}\approx 2.21$ for \SWT. Neglecting the supercurrent for \SWOT~and using $J_C^{(1)}=J_C^{(2)}\equiv J_C$, the difference in the QP density $\Delta n = \left.\left(\n^{(1)} - \n^{(2)}\right)\right|_{x=-L_1}$ at the injector is given by
\begin{align*}
 \Delta n = \left( 1 - \frac{\alpha^{(2)}}{\alpha^{(1)}} \right) \left.\n^{(1)}\right|_{x=-L_1}+\alpha^{(2)}L_S \overline{J_S}^{(2)},
\end{align*}
where the QP densities, length of the superconductor and supercurrent density are measured in units of $N_0\D$, $\xi_0$ and $\D \sigma_N/(e \xi_0)$, respectively. Plugging in all numerically determined values, $\Delta n$ acquires a value of approximately $4.8\times 10^{-3}$ for high voltages. This is in good agreement with the offset of $5.2\times10^{-3}$ in \fig~\ref{fig:CompareN_qp(V)}.

\section{Conclusion}
\label{sec:conclusion}
Normal-metal QP traps can improve the performance of superconducting devices. The superconducting proximity effect takes a central role in the evacuation process of non-equilibrium QPs. When attaching such a QP trap in close proximity to an NIS-junction, the main effects of the inverse proximity effect are a significant reduction of both the spectral gap $\Omega$ in the DOS and the $|E| = \D$ peak in the superconducting DOS. While the trapping performance arises from the latter effect, the former leads to QP poisoning due to the occupation of the new available states $\Omega < |E| < \D$. Due to Andreev reflection, which still occurs up to energies $|\Delta|$, these states contribute to the conversion from normal to supercurrent along the superconductor, which qualitatively explains the numerically observed reduction of the QP density for high injection voltages in presence of a trap. These effects need to be taken into account for finding the optimal trap position and optimizing the trapping performance. This is subject to further investigation. QP recombination and phonon emission with phonons traveling through the substrate play an important role in the poisoning~\cite{patel2017phonon}. Incorporating the phononic Green's functions in the formalism~\cite{rammer1986quantum} is beyond the current manuscript. In addition, a one-dimensional approach might not be sufficient to model extended trap geometries such as a trap array~\cite{patel2017phonon} since the kinetic properties of two-dimensional metallic proximity systems can substantially differ from those of quasi-one-dimensional structures~\cite{wilhelm1998coherent}.

\begin{acknowledgments}
We thank Pauli Virtanen, Tero T. Heikkilä and Britton Plourde for helpful discussions.
\end{acknowledgments}





\appendix*
\section{Approximate solution}
\label{app:ApproxSol}
For \SWOT~approximate solutions to the Usadel equations~\eqref{eq:spec_Usadel1}-\eqref{eq:kinetic2} can be obtained by discarding the self-consistency equation and instead using the BCS bulk value $\D$ for the order parameter. This approach neglects the supercurrent and inverse proximity effect as well as the degradation of the order parameter due to QPs and a current flow. This assumption is in agreement with numerical results. 

For the spectral quantities $\theta, \phi$, the proximity effect is neglected as well. Hence, they are given by their respective bulk solutions as well, 
\begin{align*}
 \theta &= \theta_{\text{BCS}} = \left\{ \begin{array}{ll} \frac{\pi}{2} + \frac{i}{2} \ln{\frac{1 + \epsilon}{1 - \e}} &, |\e| < 1 \\ \frac{i}{2} \ln{\frac{\e+1}{\e-1}} &, |\e| > 1 \end{array} \right. \\
 \phi &\equiv 0.
\end{align*}

Thereby, the spectral coefficients entering the kinetic equations~\eqref{eq:kinetic1}-\eqref{eq:kinetic2} are given by
\begin{align*}
 N_S &= \Theta(|\e| - 1) \frac{|\e|}{\sqrt{\e^2-1}} \\
 D_L &= \Theta(|\e| - 1) \\
 D_T &= \left\{ \begin{array}{lc} \frac{1}{1-\e^2} &, |\e|<1 \\ \frac{\e^2}{\e^2-1^2} &, |\e|>1 \end{array}\right. \\
 \mathcal{R} &= 2 \Theta(1-|\e|) \frac{1}{\sqrt{1-\e^2}} \\
 j_S &= \mathcal{T} = \mathcal{L} = 0.
\end{align*}

With $N_S$ and $D_L$ both vanishing for sub-gap energies $|\e| < 1$, the Kuprianov-Lukichev boundary condition~\cite{kuprianov1988influence} for $f_L$ is an identity equation and thus must be replaced by another appropriate boundary condition in order to obtain a unique solution. This is given by the requirement of a vanishing energy current into the superconductor, $\partial_x f_L = 0$, at the tunnel barrier for energies below the gap, $|\e| < 1$, which is due to the property of superconductors being poor heat conductors.

As the spectral coefficients do not possess a space-dependence, the kinetic equations can be solved very easily, giving
\begin{align}
 f_L(x, \e) &= \text{sign}(\e)\left[\frac{N_S(\e)}{r+L_SN_S(\e)}(x-L_2) + 1 \right]\\
 f_T(x, \e) &= \frac{1}{L_S+rN_S(\e)}(x-L_2) \label{eq:approxKinetic}
\end{align}
for $1 < |\e| < e|V|/\D$, and $f_L = \text{sign}(\e), f_T = 0$ otherwise.

Note that the leakage current vanishes exactly and thus, the spectral charge current is conserved. This is not the case for the approximate solutions of the spectral Usadel equations given in Ref.~\cite{belzig1996local}, which shows that they are qualitatively valid only in equilibrium situations.

The charge current can be approximated by
\begin{align}
 \left( \frac{\D \sigma_N}{e \xi_0} \right)^{-1} J_C &= \frac{1}{r} \int \limits_{1}^{\frac{e|V|}{\D}} \frac{N_S^2}{L/r + N_S} \d \e \\ &\approx \frac{1}{r} \int \limits_{1}^{\frac{e|V|}{\D}} N_S \d \e \\
 &= \frac{1}{r} \sqrt{\left( \frac{eV}{\D} \right)^2 - 1} \label{eq:approxJ_C}
\end{align}
for $e|V| \geq \D$, where it was used that the resistance of the superconductor in the normal state is much smaller than the resistance of the tunnel junction, i.e. $L/r \ll 1 < N_S$ for energies $\e > 1$. According to \fig~\ref{fig:CompareJ_C}, this result matches the numerically found solution very well, where supercurrent was included and the order parameter was solved self-consistently. Note also, that \eq~\eqref{eq:approxJ_C} coincides with the result given in~\cite{tinkham2004introduction}.

Within this approximation, the QP density \eq~\eqref{eq:n_qp} is given by
\begin{align}
 \frac{n_{\text{qp}}(x)}{N_0 \D} = \frac{L_2-x}{r} \int \limits_{1}^{\frac{e|V|}{\D}} N_S \left[ \left( \frac{L}{r} + \frac{1}{N_S} \right)^{-1} + \left( \frac{L}{r} + N_S \right)^{-1} \right] \d \e.
\end{align}
Note that the integral is position independent as the spectral quantities are constant in space, so that the only position dependence stems from the prefactor linear in $x$ which is due to the distribution functions. The QP density at the injector, $n_{\text{qp}}(x=-L_1)$, is plotted in \fig~\ref{fig:CompareJ_C} as a function of the applied voltage.

As the supercurrent is neglected within this approximation, the total charge current is entirely carried by normal current \eq~\eqref{eq:approxJ_C}, which is consequently constant along the superconductor. This is also evident from the position independent gradient of the QP density, as both are proportional to each other. 


\bibliography{NISN_junction}
\bibliographystyle{apsrev4-1}

\end{document}

%% file: setup3.pdf_tex
\begingroup%
  \makeatletter%
  \providecommand\color[2][]{%
    \errmessage{(Inkscape) Color is used for the text in Inkscape, but the package 'color.sty' is not loaded}%
    \renewcommand\color[2][]{}%
  }%
  \providecommand\transparent[1]{%
    \errmessage{(Inkscape) Transparency is used (non-zero) for the text in Inkscape, but the package 'transparent.sty' is not loaded}%
    \renewcommand\transparent[1]{}%
  }%
  \providecommand\rotatebox[2]{#2}%
  \newcommand*\fsize{\dimexpr\f@size pt\relax}%
  \newcommand*\lineheight[1]{\fontsize{\fsize}{#1\fsize}\selectfont}%
  \ifx\svgwidth\undefined%
    \setlength{\unitlength}{612.75859432bp}%
    \ifx\svgscale\undefined%
      \relax%
    \else%
      \setlength{\unitlength}{\unitlength * \real{\svgscale}}%
    \fi%
  \else%
    \setlength{\unitlength}{\svgwidth}%
  \fi%
  \global\let\svgwidth\undefined%
  \global\let\svgscale\undefined%
  \makeatother%
  \begin{picture}(1,0.40019941)%
    \lineheight{1}%
    \setlength\tabcolsep{0pt}%
    \put(0,0){\includegraphics[width=\unitlength,page=1]{setup3.pdf}}%
    \put(0.15591394,0.00719562){\color[rgb]{0,0,0}\makebox(0,0)[lt]{\lineheight{1.25}\smash{\begin{tabular}[t]{l}$V$\end{tabular}}}}%
    \put(0.36572079,0.07511944){\color[rgb]{0,0,0}\makebox(0,0)[lt]{\lineheight{1.25}\smash{\begin{tabular}[t]{l}$L_1$\end{tabular}}}}%
    \put(0.66456405,0.06624675){\color[rgb]{0,0,0}\makebox(0,0)[lt]{\lineheight{1.25}\smash{\begin{tabular}[t]{l}$L_2$\end{tabular}}}}%
    \put(0.3699772,0.24170214){\color[rgb]{0,0,0}\makebox(0,0)[lt]{\lineheight{1.25}\smash{\begin{tabular}[t]{l}$w$\end{tabular}}}}%
    \put(0.63563784,0.28982244){\color[rgb]{0,0,0}\makebox(0,0)[lt]{\lineheight{1.25}\smash{\begin{tabular}[t]{l}$x$\end{tabular}}}}%
    \put(0.59833859,0.33123824){\color[rgb]{0,0,0}\makebox(0,0)[lt]{\lineheight{1.25}\smash{\begin{tabular}[t]{l}$y$\end{tabular}}}}%
    \put(0.5715807,0.3630021){\color[rgb]{0,0,0}\makebox(0,0)[lt]{\lineheight{1.25}\smash{\begin{tabular}[t]{l}$z$\end{tabular}}}}%
    \put(0.84956128,0.24376046){\color[rgb]{0,0,0}\makebox(0,0)[lt]{\lineheight{1.25}\smash{\begin{tabular}[t]{l}$d_N$\end{tabular}}}}%
    \put(0.84853217,0.17354892){\color[rgb]{0,0,0}\makebox(0,0)[lt]{\lineheight{1.25}\smash{\begin{tabular}[t]{l}$d_S$\end{tabular}}}}%
  \end{picture}%
\endgroup%

%% file: InfOverlapp.pdf_tex
\begingroup%
  \makeatletter%
  \providecommand\color[2][]{%
    \errmessage{(Inkscape) Color is used for the text in Inkscape, but the package 'color.sty' is not loaded}%
    \renewcommand\color[2][]{}%
  }%
  \providecommand\transparent[1]{%
    \errmessage{(Inkscape) Transparency is used (non-zero) for the text in Inkscape, but the package 'transparent.sty' is not loaded}%
    \renewcommand\transparent[1]{}%
  }%
  \providecommand\rotatebox[2]{#2}%
  \newcommand*\fsize{\dimexpr\f@size pt\relax}%
  \newcommand*\lineheight[1]{\fontsize{\fsize}{#1\fsize}\selectfont}%
  \ifx\svgwidth\undefined%
    \setlength{\unitlength}{548.96611025bp}%
    \ifx\svgscale\undefined%
      \relax%
    \else%
      \setlength{\unitlength}{\unitlength * \real{\svgscale}}%
    \fi%
  \else%
    \setlength{\unitlength}{\svgwidth}%
  \fi%
  \global\let\svgwidth\undefined%
  \global\let\svgscale\undefined%
  \makeatother%
  \begin{picture}(1,0.29474964)%
    \lineheight{1}%
    \setlength\tabcolsep{0pt}%
    \put(0,0){\includegraphics[width=\unitlength,page=1]{InfOverlapp.pdf}}%
    \put(0.11432357,0.17624038){\color[rgb]{0,0,0}\makebox(0,0)[lt]{\lineheight{1.25}\smash{\begin{tabular}[t]{l}$d_1$\end{tabular}}}}%
    \put(0.11432357,0.06421163){\color[rgb]{0,0,0}\makebox(0,0)[lt]{\lineheight{1.25}\smash{\begin{tabular}[t]{l}$d_2$\end{tabular}}}}%
    \put(0.90263124,0.1363292){\color[rgb]{0,0,0}\makebox(0,0)[lt]{\lineheight{1.25}\smash{\begin{tabular}[t]{l}$x$\end{tabular}}}}%
    \put(0.40375849,0.17463838){\color[rgb]{0,0,0}\makebox(0,0)[lt]{\lineheight{1.25}\smash{\begin{tabular}[t]{l}$S_1,\Delta_1$\end{tabular}}}}%
    \put(0.40375849,0.06534197){\color[rgb]{0,0,0}\makebox(0,0)[lt]{\lineheight{1.25}\smash{\begin{tabular}[t]{l}$S_2,\Delta_2$\end{tabular}}}}%
    \put(0,0){\includegraphics[width=\unitlength,page=2]{InfOverlapp.pdf}}%
    \put(0.00695586,0.25322983){\color[rgb]{0,0,0}\makebox(0,0)[lt]{\lineheight{1.25}\smash{\begin{tabular}[t]{l}$z$\end{tabular}}}}%
  \end{picture}%
\endgroup%